# Unscented Auxiliary Particle Filter Implementation of the Cardinalized Probability Hypothesis Density Filter

Meysam R. Danaee

*Abstract*—**The probability hypothesis density (PHD) filter alleviates the computational expense of the optimal Bayesian multi-target filtering by approximating the intensity function of the random finite set (RFS) of targets in time. However, as a powerful decluttering algorithm, it suffers from lack of the precise estimation of the expected number of targets. The cardinalized PHD (CPHD) recursion, as a generalization of the PHD recursion, is to remedy this flaw, which jointly propagates the intensity function and the posterior cardinality distribution. While there are a few new approaches to enhance the Sequential Monte Carlo (SMC) implementation of the PHD filter, current SMC implementation for the CPHD filter suffers from poor performance in terms of accuracy of estimate. In this paper, based on the unscented transform (UT), we propose an auxiliary implementation of the CPHD filter for highly nonlinear systems. To that end, we approximate the elementary symmetric functions both with the predicted and with the update estimate of the linear functional. We subsequently demonstrate via numerical simulations that our algorithms significantly out performs both the SMC-CPHD filter and the auxiliary particle implementation of the PHD filter in difficult situations with high clutter. We also compare our proposed algorithm with its counterparts in terms of other metrics, such as run times and sensitivity to new target appearance.**

## I. INTRODUCTION

IN the context of the multi-target tracking (MTT), we have to estimate the set of unknown target states and its time-varying cardinality. Mahler provides an excellent context to cope with these difficulties. This is done by introducing the concept of finite set statistics (FISST) [1-3]. The major drawback of the FISST Bayes filter is facing manifold integrals of the multi target states in high dimensional space, which makes practical implementations of tracking system impossible. A first moment approximation to the full multi-target Bayes recursion, namely the probability hypothesis density (PHD), is introduced by Mahler [4] to propagate the posterior intensity function rather than the multi-target posterior density in time and reduce computational expense [4].

In order to emerge higher moment information into the PHD recursion, Mahler suggested a generalization of the PHD filter named the cardinalized PHD (CPHD) filter which simultaneously propagates the cardinality distribution together with the intensity function [5, 6]. There are many application areas for this powerful filtering strategy such as image tracking [7], sonar [8] , extended target tracking [9], and data fusion [10, 11].

The sequential Monte Carlo (SMC) implementation of the PHD and CPHD filter are first introduced in [2, 12, 13] using state transition density as a proposal distribution. Consequently, there are a few attempts made to enhance the SMC implementation of the PHD filter [14-16]. However, despite the supposed superiority of the CPHD over the PHD filter, to the best knowledge of the author, there is no apparently parallel approach for the auxiliary particle implementation of the CPHD filter. This is because the CPHD filter has more complex recursion compared to the PHD filter. Our goal is to make this filter efficiently implemented by exploiting the state-of-art auxiliary variable particle filtering (AVPF) algorithm and this paper is an extension to our previous work [17].

In this paper we suggest the auxiliary particle implementation of the CPHD filter which significantly outperforms the original SMC implantation of the CPHD filter proposed in [13]. To that end, we first simplify the recursion equation of the CPHD filter in a way that is suitable to apply auxiliary particle filter principle for implementation of the CPHD filter. We then show that now current samples, which approximate the intensity function and the cardinality distribution, can be drawn based on the set of auxiliary variables involving measurement indices and existing particles. We build the proposal distributions of the auxiliary variables by means of potential functions. Furthermore, we show that the elementary symmetric functions, which are the basic components of the CPHD filter, can be derived in terms of these potential functions. So, we apply UT in two steps to approximate these potential functions for both constructing the proposal distribution and computing importance weights. Although our suggested filter is expected to be computationally intensive, it would be worth using this method because, as we will show, this filter outperforms its counterparts both in cardinality estimation and localization accuracy, even in harsh situations such as in scenarios with

Meysam R. Danaee is with the EE Dept., Sharif University of Technology, Tehran, Iran and Imam Hossein Comprehensive University (IHCU), Tehran, Iran (email: mrdanaee@gmail.com).

high clutter.

The structure of this paper is as follows. The CPHD recursion is introduced in Section II. The implementation of the CPHD filter based on the idea of the auxiliary particle filter is described in Section III. The application of UT for construction of the proposal distributions is discussed in Section IV. In Section V, through a simulation scenario of MTT with various clutter levels, the performance of our proposed algorithm is evaluated in terms of different metrics and compared with counterpart algorithms. Closing remarks are given in Section VI.

## II. CARDINALIZED PHD FILTER

According to finite set statistics (FISST) framework, which is the appropriate formulation of the point process theory [18] in MTT application, the multi-target state is represented as a finite set $X = \{\mathbf{x}_1,...,\mathbf{x}_n\}$ with target state-vectors $\mathbf{x}_1,...,\mathbf{x}_n$ and random target number $n$. So the state $X$ is a random finite set $\Xi$.

Mahelr [4], analogous to constant-gain Kalman filter, developed a first-order statistical moment of the full multi-target probability density which is named the probability hypothesis density (PHD)

$$D_\Xi(\mathbf{x}) \triangleq E\big[\delta_\Xi(\mathbf{x})\big] = \int \delta_X(\mathbf{x}) f_\Xi(X) \delta X \qquad (1)$$

where $\delta_w(\mathbf{x})$ denotes the Dirac delta function concentrated at $w$.

The PHD filter serves as a powerful technique for decluttering. However, [19] argues that estimating the number of targets, the merit obtained by using the PHD filter, is subject to a large variance due to propagation of only the first-order statistical moment of the full target posterior distribution.

To that end, [5] proposed cardinalized PHD (CPHD), which propagates not only PHD but also the whole probability distribution on the number of targets. The prediction and correction steps of the CPHD filter are given by the following section.

### A. CPHD Prediction

The intensity function $D_{k|k}(\mathbf{x}_k)$ is forwarded in time by retaining survived targets (persistent targets) as well as adding newly born targets

$$D_{k+1|k}(\mathbf{x}_{k+1}) = \int_{E'} f^a(\mathbf{x}_k \mid \mathbf{x}_k) \cdot p_S^a(\mathbf{x}_k) \cdot D_{k|k}^a(\mathbf{x}_k) d\mathbf{x}_k \qquad (2)$$

where we extend the state space $E$ of the survived target with dimension $d$, $E \subset \mathbb{R}^d$ by an isolated 'source' point $\mathbf{S}$, representing a state space of newly born target, so that we may denote by $E'$ the extended space $E \cup \{\mathbf{S}\}$. Individual targets evolve independently according to the extended single-state Markov transition density

$$f^a(\mathbf{x}_{k+1} \mid \mathbf{x}_k) = \begin{cases} f(\mathbf{x}_{k+1} \mid \mathbf{x}_k) & \mathbf{x}_k \in E \\ b(\mathbf{x}_{k+1})/b[1] & \mathbf{x}_k = \mathbf{S} \end{cases} \qquad (3)$$

where $b(\mathbf{x})$ is the birth intensity function, which in this paper follows the Gaussian distribution defined as $b(\mathbf{x}) = \gamma_b \cdot \mathcal{N}(\mathbf{x}; \mathbf{m}_b, \mathbf{Q}_b)$, where $\gamma_b$ is the predicted number of newly born targets at each time-step and $\mathbf{m}_b$ is the state vector of a newly born target most likely to appear in the observation area, and $\mathbf{Q}_b$ is the uncertainty covariance matrix.[1] Furthermore, $b[h]$ is the corresponding linear functional of the function $h(\mathbf{x})$ and defined as $b[h] = \int h(\mathbf{x}) b(\mathbf{x}) d\mathbf{x}$.

Consider a target with the x and y positions $[x_k, y_k]$, corresponding velocities $[v_{x,k}, v_{y,k}]$ and the turn rate $w_k$. We assume that the target trajectories are modeled by nonlinear nearly-constant turn model [20] to represent $f(\mathbf{x}_{k+1} \mid \mathbf{x}_k)$. That model describes the evolution of the target state $\mathbf{x}_k = [x_k, v_{x,k}, y_k, v_{y,k}, w_k]'$ as

$$\mathbf{x}_{k+1} = F(w_k) \cdot \mathbf{x}_k + \mathbf{G} \cdot \varepsilon_k =$$

$$= \begin{bmatrix} 1 & \sin(w_k \cdot T)/w_k & 0 & -(1-\cos(w_k \cdot T))/w_k & 0 \\ 0 & \cos(w_k \cdot T) & 0 & -\sin(w_k \cdot T) & 0 \\ 0 & (1-\cos(w_k \cdot T))/w_k & 1 & \sin(w_k \cdot T)/w_k & 0 \\ 0 & \sin(w_k \cdot T) & 0 & \cos(w_k \cdot T) & 0 \\ 0 & 0 & 0 & 0 & 1 \end{bmatrix} \cdot \mathbf{x}_k$$

$$+ \begin{bmatrix} T^2/2 & 0 & 0 \\ T & 0 & 0 \\ 0 & T^2/2 & 0 \\ 0 & T & 0 \\ 0 & 0 & 1 \end{bmatrix} \cdot \varepsilon_k \qquad (4)$$

where the process noise $\varepsilon_k$ has a multivariate normal distribution defined as $\mathcal{N}(\cdot; \mathbf{0}_{3\times1}, \mathbf{Q})$ with $\mathbf{Q} = diag(\sigma_{\tilde{x}}^2, \sigma_{\tilde{x}}^2, \sigma_w^2)$ and $T$ is the sampling time. As a result, we may write $f(\mathbf{x}_{k+1} \mid \mathbf{x}_k) = \mathcal{N}(\mathbf{x}_{k+1}; F(w_k) \cdot \mathbf{x}_k, \mathbf{G} \cdot \mathbf{Q} \cdot \mathbf{G}')$.

The extended probability of survival for a target with state $\mathbf{x}_k$ at time-step $k$ is given by

$$p_S^a(\mathbf{x}_k) = \begin{cases} p_S(\mathbf{x}_k) & \mathbf{x}_k \in E \\ 1 & \mathbf{x}_k = \mathbf{S} \end{cases} \qquad (5)$$

where $p_S(\mathbf{x}_k)$ is the survival probability of the persistent target with state $\mathbf{x}_k$. We also denote by $D_{k|k}^a(\mathbf{x}_k)$ the extended intensity function defined as

$$D_{k|k}^a(\mathbf{x}_k) = b[1] \delta_S(\mathbf{x}_k) + D_{k|k}(\mathbf{x}_k). \qquad (6)$$

The prediction cardinality distribution is given by

$$p_{k+1|k}(n) = \sum_{i=0}^{n} p_B(n-i) \cdot \frac{1}{i!} \cdot G^{(i)}(1-s[p_S]) \cdot s[p_S]^i \qquad (7)$$

where $p_B(n)$ is the cardinality distribution of newly born targets, $G(x) = G_{k|k}(x)$ is the probability generating function (PGF) of the cardinality distribution of $f_{k|k}(X \mid Z^{(k)})$, $s[h] = \int h(\mathbf{x}) s(\mathbf{x}) \, d\mathbf{x}$ where $s(\mathbf{x}) = N_{k|k}^{-1} . D_{k|k}(\mathbf{x})$, and $N_{k|k} = \int D_{k|k}(\mathbf{x} \mid Z^{(k)}) \, d\mathbf{x}$. Note that $G^{(i)}(x)$ is the $i$th derivative of $G(x)$.

### B. CPHD Correction

At time-step $k+1$ we receive measurements which comprise a finite set $Z_{k+1} = \{\mathbf{z}_1, \ldots, \mathbf{z}_m\}$. Those measurements belonging to the true targets are generated based on the range and bearing measurement model

$$\mathbf{z} = \left[ \sqrt{x^2 + y^2} \quad \arctan\left(\frac{y}{x}\right) \right]' + \begin{bmatrix} v_r \\ v_\theta \end{bmatrix}$$

$$\mathbf{R} = E\left\{ \begin{bmatrix} v_r \\ v_\theta \end{bmatrix} \begin{bmatrix} v_r & v_\theta \end{bmatrix} \right\} = diag\left(\sigma_r^2, \sigma_\theta^2\right) \tag{8}$$

where $v_r$ and $v_\theta$ are zero mean Gaussian measurement noises of variance $\sigma_r^2$ and $\sigma_\theta^2$ respectively. Furthermore, we adopt the notation $Z^{(k+1)} : Z_1, \ldots, Z_{k+1}$. The measurements which do not belong to true targets are false alarm events which we assume that they are Poisson processes with spatial distribution $c(\mathbf{z})$ and average number of clutter points $\lambda$.

Each target is detected with the probability of detection $p_D(\mathbf{x})$.

The correction step of the CPHD filter is given by

$$D_{k+1|k+1}(\mathbf{x}) \cong L_{Z_{k+1}}(\mathbf{x}) . D_{k+1|k}(\mathbf{x}) \tag{9}$$

where we denote

$$L_Z(\mathbf{x}) \triangleq \frac{\Upsilon^1(Z)}{\Upsilon^0(Z)} . q_D(\mathbf{x})$$
$$+ p_D(\mathbf{x}) . \sum_{p=1}^{|Z|} \left\{ \frac{L_{\mathbf{z}_p}(\mathbf{x})}{c(\mathbf{z}_p)} . \frac{\Upsilon^1(Z - \{\mathbf{z}_p\})}{\Upsilon^0(Z)} \right\}, \tag{10}$$

where $L_{\mathbf{z}_p}(\mathbf{x}) = f(\mathbf{z}_p \mid \mathbf{x})$ is the measurement likelihood function. We denote $q_D(\mathbf{x}) = 1 - p_D(\mathbf{x})$. In addition, we define the function $\Upsilon^u(Z)$ as

$$\Upsilon^u(Z) = \sum_{j=0}^{|Z|} \frac{(|Z|-j)!}{(N_{k+1|k})^{j+u}} . p_k(|Z|-j) . \sigma_j(Z)$$
$$. \sum_{n=j+u}^{\infty} P_{j+u}^n . q_D^{n-(j+u)} . p(n), \tag{11}$$

where $p_k(n)$ is the cardinality distribution of false alarms, $p(n) = p_{k+1|k}(n)$ is the predicted cardinality distribution at time-step $k+1$, $P_\ell^n$ is the k-permutations of n, and $\sigma_i(Z)$ is given by

$$\sigma_i(Z) \triangleq \sigma_{m,i}\left( \frac{D_{k+1|k}[p_D L_{\mathbf{z}_1}]}{c(\mathbf{z}_1)}, \ldots, \frac{D_{k+1|k}[p_D L_{\mathbf{z}_m}]}{c(\mathbf{z}_m)} \right), \tag{12}$$

where the elementary symmetric function $\sigma_{m,i}(y_1, \ldots, y_m)$ of

degree $i$ in $y_1, \ldots, y_m$ is given by

$$\sigma_{m,i}(y_1, \ldots, y_m) = \sum_{S \subseteq U \atop |S|=i} \prod_{j \in S} y_j, \tag{13}$$

where the degree zero is given by convention as

$$\sigma_{m,0}(y_1, \ldots, y_m) = 1, \tag{14}$$

and we denote by $D_{k+1|k}[h]$

$$D_{k+1|k}[h] = \int D_{k+1|k}(\mathbf{x}) h(\mathbf{x}) d\mathbf{x}. \tag{15}$$

We assume that $p_D(\mathbf{x})$ is independent of state-vector $\mathbf{x}$ for all $\mathbf{x}$ and we may, therefore, write $s[q_D] = q_D$.

Consequently, the cardinality distribution is updated in terms of the predicted cardinality distribution as

$$p_{k+1|k+1}(n) = \frac{p_{k+1|k}(n)}{\Upsilon^0(Z_{k+1})}$$
$$\left( \begin{array}{c} \sum_{j=0}^{|Z_{k+1}|} (|Z_{k+1}|-j)! . p_k(|Z_{k+1}|-i) . \sigma_j(Z_{k+1}) \\ . P_j^n . \frac{q_D^{n-j}}{(N_{k+1|k})^j} \end{array} \right). \tag{16}$$

### III. AUXILIARY PARTICLE FILTER IMPLEMENTATION

To convey the idea behind our proposed approach, we first consider a discrete approximation of the following integral

$$\overline{\varphi} = \int_E \varphi(\mathbf{x}_{k+1}) D_{k+1|k+1}(\mathbf{x}_{k+1}) d\mathbf{x}_{k+1} \tag{17}$$

by using samples generated from a proposal distribution, where $\varphi$ is a test function. Instead of forwarding all existing particles with an inefficient proposal distribution such as $f^\alpha(\mathbf{x}_{k+1} \mid \mathbf{x}_k)$, we may construct an efficient proposal distribution. We achieve this goal by using auxiliary particle filter principle and sampling on a higher dimensional space in the hope that it will increase estimation accuracy.

The principle of the Auxiliary Particle Filter (APF) [21], which is using auxiliary random variables may be applied in the target tracking contexts in which the number of targets are not already known and can be changed during time. The application of the auxiliary random variable can be extended for detection and tracking a target within raw measurements [22] or multiple target tracking using the PHD filter with time varying number of targets [15].

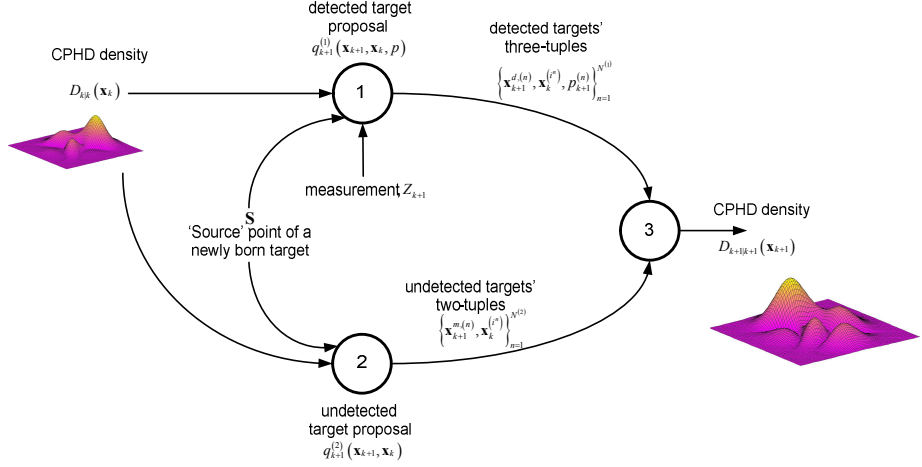

**Figure 1. Auxiliary CPHD algorithm implementation.** Two proposal distributions $q_{k+1}^{(1)}\left(\mathbf{x}_{k+1}, \mathbf{x}_k, p\right)$ (for detected targets) and $q_{k+1}^{(2)}\left(\mathbf{x}_{k+1}, \mathbf{x}_k\right)$ (for undetected targets) generate samples to approximate $D_{k+1|k+1}\left(\mathbf{x}_{k+1}\right)$. Based on the source point of a newly born target, existing particles $\{\mathbf{x}_k^n\}_{n=1}^{N_p}$ which are approximating $D_{k|k}\left(\mathbf{x}_k\right)$, and current measurement set $Z_{k+1}$, $N^{(1)}$ three-tuples $\{\mathbf{x}_{k+1}^{d,(n)}, \mathbf{x}_k^{(i^n)}, p_{k+1}^{(n)}\}_{n=1}^{N^{(1)}}$ are sampled while the super script d in $\mathbf{x}_{k+1}^{d,(n)}$ denotes samples belong to detected targets. For undetected targets, $q_{k+1}^{(2)}\left(\mathbf{x}_{k+1}, \mathbf{x}_k\right)$ draws $N^{(2)}$ two-tuples $\{\mathbf{x}_{k+1}^{m,(n)}, \mathbf{x}_k^{(i^n)}\}_{n=1}^{N^{(2)}}$ where while the super script m in $\mathbf{x}_{k+1}^{m,(n)}$ denotes samples belong to undetected targets. We take the union of these $N_p = N^{(1)} + N^{(2)}$ tuples with

We explore a situation in which targets are detected by probability density $q_{k+1}^{(1)}\left(\mathbf{x}_{k+1}, \mathbf{x}_k, p\right)$ which defined on spaces $E \times E' \times |Z_{k+1}|$. In this situation, auxiliary variables include existing particles $\{\mathbf{x}_k^n\}_{n=1}^{N_p}$ and indices of current measurements $p \in \{1, ..., |Z_{k+1}|\}$. Consequently, sampling on a higher dimensional space is done by first picking up a measurement which can be regarded as an observation more probably generated from a true target. Secondly, we select an existing particle on the basis of how well it describes the picked up measurement. We also explore a situation of undetected targets by proposal distribution $q_{k+1}^{(2)}(\mathbf{x}_{k+1}, \mathbf{x}_k)$ defined on spaces $E \times E'$ by using existing particles $\{\mathbf{x}_k^n\}_{n=1}^{N_p}$ as auxiliary variables.

The fundamental flow of the Auxiliary CPHD (ACPHD) algorithm is shown in Fig. 1.

We substitute $D_{k+1|k+1}\left(\mathbf{x}_{k+1}\right)$ appeared in (17) with its comprising terms defined in (2) and (9), and then we use the proposal distributions $q_{k+1}^{(1)}\left(\mathbf{x}_{k+1}, \mathbf{x}_k, p\right)$ and $q_{k+1}^{(2)}\left(\mathbf{x}_{k+1}, \mathbf{x}_k\right)$ in order to change the integrand in (17) to

$$
\begin{aligned}
\bar{\varphi} = &\sum_{p=1}^{|Z_{k+1}|} \int_{E'} \int_E \varphi\left(\mathbf{x}_{k+1}\right). \frac{p_D\left(\mathbf{x}_{k+1}\right).L_{\mathbf{z}_p}\left(\mathbf{x}_{k+1}\right)}{c\left(\mathbf{z}_p\right)} \\
&\times \frac{\Upsilon^1\left(Z_{k+1} - \{\mathbf{z}_p\}\right)}{\Upsilon^0\left(Z_{k+1}\right)} \\
&\times \frac{f^a\left(\mathbf{x}_{k+1} \mid \mathbf{x}_k\right). p_S^g\left(\mathbf{x}_k\right). D_{k|k}^a\left(\mathbf{x}_k\right)}{q_{k+1}^{(1)}\left(\mathbf{x}_{k+1}, \mathbf{x}_k, p\right)}
\end{aligned}
\tag{18}
$$

$$
\begin{aligned}
&\times q_{k+1}^{(1)}\left(\mathbf{x}_{k+1}, \mathbf{x}_k, p\right) d\mathbf{x}_k d\mathbf{x}_{k+1} \\
&+ \int_{E'} \int_E \varphi\left(\mathbf{x}_{k+1}\right).\left(1 - p_D\left(\mathbf{x}_{k+1}\right)\right). \frac{\Upsilon^1\left(Z_{k+1}\right)}{\Upsilon^0\left(Z_{k+1}\right)} \\
&\times \frac{f^a\left(\mathbf{x}_{k+1} \mid \mathbf{x}_k\right). p_S^g\left(\mathbf{x}_k\right). D_{k|k}^a\left(\mathbf{x}_k\right)}{q_{k+1}^{(2)}\left(\mathbf{x}_{k+1}, \mathbf{x}_k\right)} \\
&\times q_{k+1}^{(2)}\left(\mathbf{x}_{k+1}, \mathbf{x}_k\right) d\mathbf{x}_k d\mathbf{x}_{k+1}.
\end{aligned}
$$

According to the Bayes' rule, we can decompose the joint proposal distributions as

$$
\begin{aligned}
q_{k+1}^{(1)}\left(\mathbf{x}_{k+1}, \mathbf{x}_k, p\right) = &q_{k+1}^{(1)}\left(\mathbf{x}_{k+1} \mid \mathbf{x}_k, p\right) \\
&\times q_{k+1}^{(1)}\left(\mathbf{x}_k \mid p\right). q_{k+1}^{(1)}(p),
\end{aligned}
\tag{19}
$$

and

$$
q_{k+1}^{(2)}\left(\mathbf{x}_{k+1}, \mathbf{x}_k\right) = q_{k+1}^{(2)}\left(\mathbf{x}_{k+1} \mid \mathbf{x}_k\right). q_{k+1}^{(2)}\left(\mathbf{x}_k\right).
\tag{20}
$$

We can also trivially prove that the minimum variance of importance weights is direct consequence of the following choices of decompositions

$$
\begin{aligned}
q_{k+1}^{(1)}(p) \propto &\frac{\Upsilon^1\left(Z_{k+1} - \{\mathbf{z}_p\}\right)}{c\left(\mathbf{z}_p\right)} \\
&\times \int_{E'} \mathcal{V}_{k+1,p}^{(1)}\left(\mathbf{x}_k\right). D_{k|k}^a\left(\mathbf{x}_k\right) d\mathbf{x}_k, \quad p = 1, ..., |Z_{k+1}|,
\end{aligned}
\tag{21}
$$

$$
q_{k+1}^{(1)}\left(\mathbf{x}_k \mid p\right) = \frac{\mathcal{V}_{k+1,p}^{(1)}\left(\mathbf{x}_k\right). D_{k|k}^a\left(\mathbf{x}_k\right)}{\int_{E'} \mathcal{V}_{k+1,p}^{(1)}\left(\mathbf{x}_k\right). D_{k|k}^a\left(\mathbf{x}_k\right) d\mathbf{x}_k},
\tag{22}
$$

$$
q_{k+1}^{(2)}\left(\mathbf{x}_k\right) = \frac{\mathcal{V}_{k+1}^{(2)}\left(\mathbf{x}_k\right). D_{k|k}^a\left(\mathbf{x}_k\right)}{\int_{E'} \mathcal{V}_{k+1}^{(2)}\left(\mathbf{x}_k\right). D_{k|k}^a\left(\mathbf{x}_k\right) d\mathbf{x}_k},
\tag{23}
$$

where $\mathcal{V}_{k+1,p}^{(1)}(\mathbf{x}_k)$ and $\mathcal{V}_{k+1}^{(2)}(\mathbf{x}_k)$ are bounded potential functions for detected and undetected targets respectively, and defined as follows

$$\mathcal{V}_{k+1,p}^{(1)}(\mathbf{x}_k) = \int_E p_D(\mathbf{x}_{k+1}) \cdot L_{z_p}(\mathbf{x}_{k+1}) \cdot \qquad (24)$$
$$\times f^a(\mathbf{x}_{k+1} \mid \mathbf{x}_k) \cdot p_S^a(\mathbf{x}_k) d\mathbf{x}_{k+1},$$

$$\mathcal{V}_{k+1}^{(2)}(\mathbf{x}_k) = \int_E (1 - p_D(\mathbf{x}_{k+1})) \qquad (25)$$
$$\times f^a(\mathbf{x}_{k+1} \mid \mathbf{x}_k) \cdot p_S^a(\mathbf{x}_k) d\mathbf{x}_{k+1}.$$

We apply the potential function $\mathcal{V}_{k+1,p}^{(1)}(\mathbf{x}_k)$ to enforce $p$th current measurement's fitness on the selection of $\mathbf{x}_k$, while $\mathcal{V}_{k+1}^{(2)}(\mathbf{x}_k)$ can determine how likely the existing particle $\mathbf{x}_k$ is to be undetected at the time-step $k+1$. Estimations of these potential functions by both UT method (in prediction mode) and drawn samples (in update mode) are the key elements of our algorithms. This is due to the fact that, as we will show later on in the paper, other parameters can be readily computed based on these potential functions.

Now suppose that, we have a discrete approximation of the CPHD filter at time-step $k$ with a limited number of particles and related weights $\{\mathbf{x}_k^{(n)}, w_k^{(n)}\}_{n=1}^{N_p}$

$$\tilde{D}_{k|k}(\mathbf{x}_k) = \sum_{i=1}^{N_p} w_k^{(i)} \cdot \delta_{\mathbf{x}_k^{(i)}}(\mathbf{x}_k), \qquad (26)$$

and therefore

$$\tilde{D}_{k|k}^a(\mathbf{x}_k) = \tilde{D}_{k|k}(\mathbf{x}_k) + b[1] \delta_S(\mathbf{x}_k). \qquad (27)$$

These approximations become exact when the number of particles grows to infinity. Based on the discrete approximation in (27), the equations from (21) to (23) are adjusted as

$$q_{k+1}^{(1)}(p) \propto \frac{\Upsilon^1 \left( Z_{k+1} - \{\mathbf{z}_p\} \right)}{c(\mathbf{z}_p)} \qquad (28)$$
$$\times \left( \sum_{i=1}^{N_p} \mathcal{V}_{k+1,p}^{(1)}(\mathbf{x}_k^{(i)}) \cdot w_k^{(i)} + \mathcal{V}_{k+1,p}^{(1)}(\mathbf{S}) \cdot b[1] \right),$$

$$q_{k+1}^{(1)}(\mathbf{x}_k \mid p) = \frac{\sum_{i=1}^{N_p} \mathcal{V}_{k+1,p}^{(1)}(\mathbf{x}_k^{(i)}) \cdot w_k^{(i)} \cdot \delta_{\mathbf{x}_k^{(i)}}(\mathbf{x}_k)}{\sum_{i=1}^{N_p} \mathcal{V}_{k+1,p}^{(1)}(\mathbf{x}_k^{(i)}) \cdot w_k^{(i)} + \mathcal{V}_{k+1,p}^{(1)}(\mathbf{S}) \cdot b[1]} \qquad (29)$$
$$+ \frac{\mathcal{V}_{k+1,p}^{(1)}(\mathbf{S}) \cdot b[1] \cdot \delta_S(\mathbf{x}_k)}{\sum_{i=1}^{N_p} \mathcal{V}_{k+1,p}^{(1)}(\mathbf{x}_k^{(i)}) \cdot w_k^{(i)} + \mathcal{V}_{k+1,p}^{(1)}(\mathbf{S}) \cdot b[1]},$$

$$q_{k+1}^{(2)}(\mathbf{x}_k) = \frac{\sum_{i=1}^{N_p} \mathcal{V}_{k+1}^{(2)}(\mathbf{x}_k^{(i)}) \cdot w_k^{(i)} \cdot \delta_{\mathbf{x}_k^{(i)}}(\mathbf{x}_k)}{\sum_{i=1}^{N_p} \mathcal{V}_{k+1}^{(2)}(\mathbf{x}_k^{(i)}) \cdot w_k^{(i)} + \mathcal{V}_{k+1}^{(2)}(\mathbf{S}) \cdot b[1]} \qquad (30)$$
$$+ \frac{\mathcal{V}_{k+1}^{(2)}(\mathbf{S}) \cdot b[1] \cdot \delta_S(\mathbf{x}_k)}{\sum_{i=1}^{N_p} \mathcal{V}_{k+1}^{(2)}(\mathbf{x}_k^{(i)}) \cdot w_k^{(i)} + \mathcal{V}_{k+1}^{(2)}(\mathbf{S}) \cdot b[1]}.$$

## IV. APPLICATION OF UT TO AUXILIARY CPHD FILTER

In order to build the proposal distributions for sampling auxiliary variables, we have to determines integrals (24) and (25). However, it is generally impossible because new samples $\mathbf{x}_{k+1}$ are not yet available. Here, we can use the unscented transform to approximate the values of the potential functions based on the existing particles $\{\mathbf{x}_k^n\}_{n=1}^{N_p}$ rather than resorting to unknown samples $\mathbf{x}_{k+1}$.

One of the benefits of applying UT is that its byproducts can be used to design Gaussian proposal distributions $q_{k+1}^{(1)}(\mathbf{x}_{k+1} \mid \mathbf{x}_k, p)$, $p = 1, \dots, |Z_{k+1}|$.

For each sample $\mathbf{x}_k^{(i)}$ we keep track of a posterior covariance matrix $\mathbf{P}_k^{(i)}$, so that we form augmented state vector, $\bar{\mathbf{x}}_k^{a,(i)}$, and covariance, $\mathbf{P}_k^{a,(i)}$, with zero mean noises $\boldsymbol{\varepsilon}_k$ and $\boldsymbol{\xi}_k$ as

$$\bar{\mathbf{x}}_k^{a,(i)} = \begin{bmatrix} \mathbf{x}_k^{(i)} & \mathbf{0}_{1\times3} & \mathbf{0}_{1\times2} \end{bmatrix}', \mathbf{P}_k^{a,(i)} = \begin{bmatrix} \mathbf{P}_k^{(i)} & \mathbf{0}_{5\times3} & \mathbf{0}_{5\times2} \\ \mathbf{0}_{3\times5} & \mathbf{Q} & \mathbf{0}_{3\times2} \\ \mathbf{0}_{2\times5} & \mathbf{0}_{2\times3} & \mathbf{R} \end{bmatrix} \qquad (31)$$
$$, \quad i = 1, \dots, N_p,$$

where $\mathbf{Q}$ and $\mathbf{R}$ are defined in Section II. For the newly born target sample $\mathbf{x}_k^{(N_p+1)} = \mathbf{S}$, we should define an augmented state vector and a covariance with different dimensions as follow

$$\bar{\mathbf{x}}_k^{a,(N_p+1)} = \begin{bmatrix} \mathbf{m}_b & \mathbf{0}_{1\times5} & \mathbf{0}_{1\times2} \end{bmatrix}',$$
$$\mathbf{P}_k^{a,(N_p+1)} = \begin{bmatrix} \sigma_b^2 \cdot \mathbf{I}_{5\times5} & \mathbf{0}_{5\times5} & \mathbf{0}_{5\times2} \\ \mathbf{0}_{5\times5} & \mathbf{Q}_b & \mathbf{0}_{5\times2} \\ \mathbf{0}_{2\times5} & \mathbf{0}_{2\times5} & \mathbf{R} \end{bmatrix}, \qquad (32)$$

where $\mathbf{m}_b$ and $\mathbf{Q}_b$ are mean and covariance of the birth intensity function and $\sigma_b^2 \cdot \mathbf{I}_{5\times5}$ is a covariance matrix that expresses the uncertainty about the mean of the distribution of newly born targets, so that $\sigma_b$ could be a very small positive value.

We build a matrix $\boldsymbol{\chi}_k^{a,(i)}$ of $2L+1$ UT sigma points (vectors) to capture statistics of the stochastic process of target motion defined via the nonlinear transformation in (4) as

$$\boldsymbol{\chi}_k^{a,(i)} = \begin{bmatrix} \bar{\mathbf{x}}_k^{a,(i)} & \bar{\mathbf{x}}_k^{a,(i)} \pm \sqrt{(L+\lambda)\mathbf{P}_k^{a,(i)}} \end{bmatrix}_{L\times(2L+1)} \qquad (33)$$

where $L$ is the dimension of the augmented state and $\gamma = \alpha^2(L+\kappa) - L$ is a scaling parameter with a secondary scaling parameters $\kappa$ and $\alpha$. Relating UT weights are defined according to

$$W_0^{(m)} = \lambda / (L+\lambda), W_0^{(c)} = \lambda / (L+\lambda) + (1-\alpha^2+\beta) \qquad (34)$$
$$W_j^{(m)} = W_j^{(c)} = 0.5 / (L+\lambda), \quad j = 1, \dots, 2L$$

where $\beta$ is a constant.

Going for the time update, we form the $j$th predicted sigma point state, $\boldsymbol{\chi}_{k+1|k}^{x,(i)}(:,j)_{5\times1}$ as

$$\boldsymbol{\chi}_{k+1|k}^{x,(i)}(:,j) = \qquad (35)$$

$$
\begin{cases}
F\left(\boldsymbol{\chi}_k^{a,(i)}(5,j)\right) \cdot \boldsymbol{\chi}_k^{a,(i)}(1:5,j) \\
\quad + \mathbf{G} \cdot \boldsymbol{\chi}_k^{a,(i)}(6:8,j), \\
\quad j = 0,\ldots,2L, i = 1,\ldots,N_p \\
\boldsymbol{\chi}_k^{a,(i)}(1:5,j) + \boldsymbol{\chi}_k^{a,(i)}(6:10,j), \\
\quad j = 0,\ldots,2L, i = N_p + 1
\end{cases}
$$

where we denote by $\boldsymbol{\chi}(:,j)$ the $j$th column of $\boldsymbol{\chi}$ and by $\boldsymbol{\chi}(a:b,j)$ the jth column of $\boldsymbol{\chi}$ where the number of its rows goes from a to b. Note that $\boldsymbol{\chi}(5,j)$ particularly denotes the turn rate of $\boldsymbol{\chi}(:,j)$. The predicted sigma point states of (35) play a key role in approximating the potential functions. We denote by $\hat{\mathcal{V}}_{k+1,p}^{(1)}(\mathbf{x}_k^{(i)})$ the predicted estimate of the potential function $\mathcal{V}_{k+1,p}^{(1)}(\mathbf{x}_k^{(i)})$ which is computed as

$$
\hat{\mathcal{V}}_{k+1,p}^{(1)}\left(\mathbf{x}_k^{(i)}\right) = \sum_{j=0}^{2L} p_S^g\left(\boldsymbol{\chi}_{k+1|k}^{x,(i)}(:,j)\right) \cdot p_D\left(\boldsymbol{\chi}_{k+1|k}^{x,(i)}(:,j)\right) \\
\times L_{z_p}\left(\boldsymbol{\chi}_{k+1|k}^{x,(i)}(:,j)\right) \cdot W_j^{(m)}, \quad i = 1,\ldots,N_p + 1
\tag{36}
$$

The predicted estimate of potential function for the case of undetected targets, $\hat{\mathcal{V}}_{k+1}^{(2)}(\mathbf{x}_k^{(i)})$, can be computed analytically if the detection probability is independent of a target state (and in this paper we assume the same property for all states). As a result, (25) is simplified to the following form

$$
\mathcal{V}_{k+1}^{(2)}\left(\mathbf{x}_k\right) = \begin{cases} (1 - p_D) \cdot p_S & \mathbf{x}_k \neq \mathbf{S} \\ 1 - p_D & \mathbf{x}_k = \mathbf{S} \end{cases}
\tag{37}
$$

and thus there is no need to apply UT anymore.

Similar to the previous notation for $\hat{\mathcal{V}}_{k+1,p}^{(1)}(\mathbf{x}_k^{(i)})$, we denote by $\hat{D}_{k+1|k}[p_D L_{z_p}]$ the predicted estimate of the linear functional $D_{k+1|k}[p_D L_{z_p}]$, where we replace $\mathcal{V}_{k+1,p}^{(1)}(\mathbf{x}_k^{(i)})$ with $\hat{\mathcal{V}}_{k+1,p}^{(1)}(\mathbf{x}_k^{(i)})$

$$
\hat{D}_{k+1|k}[p_D L_{z_p}] = \sum_{i=1}^{N_p} \hat{\mathcal{V}}_{k+1,p}^{(1)}\left(\mathbf{x}_k^{(i)}\right) \cdot w_k^i + \hat{\mathcal{V}}_{k+1,p}^{(1)}(\mathbf{S}) \cdot b[1]
\tag{38}
$$

Now it is possible to construct the proposal distributions for sampling auxiliary variables of detected targets

$$
q_{k+1}^{(1)}\left(p_{k+1}\right) = \cfrac{\cfrac{\hat{\Upsilon}^1\left(Z_{k+1} - \{\mathbf{z}_{p_{k+1}}\}\right)}{c\left(\mathbf{z}_{p_{k+1}}\right)}}{\sum_{p=1}^{|Z_{k+1}|} \cfrac{\hat{\Upsilon}^1\left(Z_{k+1} - \{\mathbf{z}_p\}\right)}{c\left(\mathbf{z}_p\right)}}
$$
$$
\times \frac{\sum_{i=1}^{N_p+1} D_{k|k}^g\left(\mathbf{x}_k^{(i)}\right) \cdot \hat{\mathcal{V}}_{k+1,p_{k+1}}^{(1)}\left(\mathbf{x}_k^{(i)}\right)}{\sum_{i=1}^{N_p+1} D_{k|k}^g\left(\mathbf{x}_k^{(i)}\right) \cdot \hat{\mathcal{V}}_{k+1,p}^{(1)}\left(\mathbf{x}_k^{(i)}\right)},
$$
$$
, p_{k+1} \in \{1,\ldots,|Z_{k+1}|\}
\tag{39}
$$
$$
q_{k+1}^{(1)}\left(\mathbf{x}_k^{(n)} \mid p_{k+1}^{(n)}\right)
$$
$$
= \frac{D_{k|k}^g\left(\mathbf{x}_k^{(n)}\right) \cdot \hat{\mathcal{V}}_{k+1,p_{k+1}^{(n)}}^{(1)}\left(\mathbf{x}_k^{(n)}\right)}{\sum_{i=1}^{N_p+1} D_{k|k}^g\left(\mathbf{x}_k^{(i)}\right) \cdot \hat{\mathcal{V}}_{k+1,p_{k+1}^{(n)}}^{(1)}\left(\mathbf{x}_k^{(i)}\right)},
$$
$$
, n = 1,\ldots,N_p + 1
$$

where $\hat{\Upsilon}^1(Z_{k+1} - \{\mathbf{z}_p\})$ is defined in the same way as

$\Upsilon^1(Z_{k+1} - \{\mathbf{z}_p\})$, expect that $D_{k+1|k}[p_D L_{z_p}]$ is replaced by the predicted estimates $\hat{D}_{k+1|k}[p_D L_{z_p}]$ for $p = 1,\ldots,|Z_{k+1}|$. The function $\Upsilon^1(Z_{k+1} - \{\mathbf{z}_p\})$ is also defined in terms of the bounded potential functions. To see that, let's look at (11) which shows that $\Upsilon^1(Z_{k+1} - \{\mathbf{z}_p\})$ is a function of $\sigma_j(Z_{k+1} - \{\mathbf{z}_p\})$ for $j = 1,\ldots,|Z_{k+1}| - 1$, $p = 1,\ldots,|Z_{k+1}|$. These elementary symmetric functions, in turn, need the quantity of $D_{k+1|k}[p_D L_{z_p}]$, $p = 1,\ldots,|Z_{k+1}|$ which are the corresponding linear functional of the predicted intensity function, $D_{k+1|k}(\mathbf{x}_{k+1})$, and defined as

$$
D_{k+1|k}[p_D L_{z_p}]
$$
$$
= \int_E p_D(\mathbf{x}_{k+1}) . L_{z_p}(\mathbf{x}_{k+1}) \cdot D_{k+1|k}(\mathbf{x}_{k+1}) d\mathbf{x}_{k+1},
\tag{40}
$$

Now let the respective sample-based approximation of $D_{k+1|k}(\mathbf{x}_{k+1})$ be

$$
\tilde{D}_{k+1|k}(\mathbf{x}_{k+1})
$$
$$
= b(\mathbf{x}_{k+1}) + \sum_{i=1}^{N_p} w_k^i \cdot p_S\left(\mathbf{x}_k^{(i)}\right) \cdot f\left(\mathbf{x}_{k+1} \mid \mathbf{x}_k^{(i)}\right),
\tag{41}
$$

then we can trivially approximate $D_{k+1|k}[p_D L_{z_p}]$ as follows

$$
\tilde{D}_{k+1|k}[p_D L_{z_p}] = \sum_{i=1}^{N_p} \mathcal{V}_{k+1,p}^{(1)}\left(\mathbf{x}_k^{(i)}\right) \cdot w_k^i + \mathcal{V}_{k+1,p}^{(1)}(\mathbf{S}) \cdot b[1]
\tag{42}
$$

where $\tilde{D}_{k+1|k}[p_D L_{z_p}] \xrightarrow{N_p \to \infty} D_{k+1|k}[p_D L_{z_p}]$. We henceforth use the terms $D_{k+1|k}[p_D L_{z_p}]$ and $\tilde{D}_{k+1|k}[p_D L_{z_p}]$ interchangeably to avoid the unnecessary diversity of notation for syntactic definitions.

Fig. 2 shows the schematic illustration of the fundamental components of the proposal distributions $q_{k+1}^{(1)}(\mathbf{x}_k^{(i^n)} \mid p_{k+1}^{(n)})$ and $q_{k+1}^{(1)}(p_{k+1}^{(n)})$.

As we mentioned earlier, there is no need to apply UT anymore for undetected targets. Consequently, the proposal distribution for sampling $\{\mathbf{x}_k\}_{n=1}^{N^{(2)}}$ is given by

$$
q_{k+1}^{(2)}\left(\mathbf{x}_k\right) = \frac{p_S \cdot \sum_{i=1}^{N_p} w_k^{(i)} \cdot \delta_{\mathbf{x}_k^{(i)}}\left(\mathbf{x}_k\right) + b[1] \cdot \delta_S\left(\mathbf{x}_k\right)}{\sum_{i=1}^{N_p} p_S \cdot w_k^{(i)} + b[1]}
\tag{43}
$$

The last step required to obtain detected target three-tuples $\{\mathbf{x}_{k+1}^{d,(n)}, \mathbf{x}_k^{(i^n)}, p_{k+1}^{(n)}\}_{n=1}^{N^{(1)}}$ and undetected target two-tuples $\{\mathbf{x}_{k+1}^{m,(n)}, \mathbf{x}_k^{(i^n)}\}_{n=1}^{N^{(2)}}$ is the construction of two proposal distributions $q_{k+1}^{(1)}(\mathbf{x}_{k+1} \mid \mathbf{x}_k, p)$ and $q_{k+1}^{(2)}(\mathbf{x}_{k+1} \mid \mathbf{x}_k)$.

First, we concentrate on $q_{k+1}^{(1)}(\mathbf{x}_{k+1} \mid \mathbf{x}_k, p)$. Let us consider the $n$th pair of auxiliary variables $\mathbf{x}_k^{(i^n)}$ and $p_{k+1}^{(n)}$ which are selected according to (39), as a part of three-tuple $\{\mathbf{x}_{k+1}^{d,(n)}, \mathbf{x}_k^{(i^n)}, p_{k+1}^{(n)}\}_{n=1}^{N^{(1)}}$. Hereafter, we drop the index n for notational ease.

The time update for the given pair of auxiliary variables $\mathbf{x}_k^{(i)}$ and $p_{k+1}$ using unscented transform includes the following steps

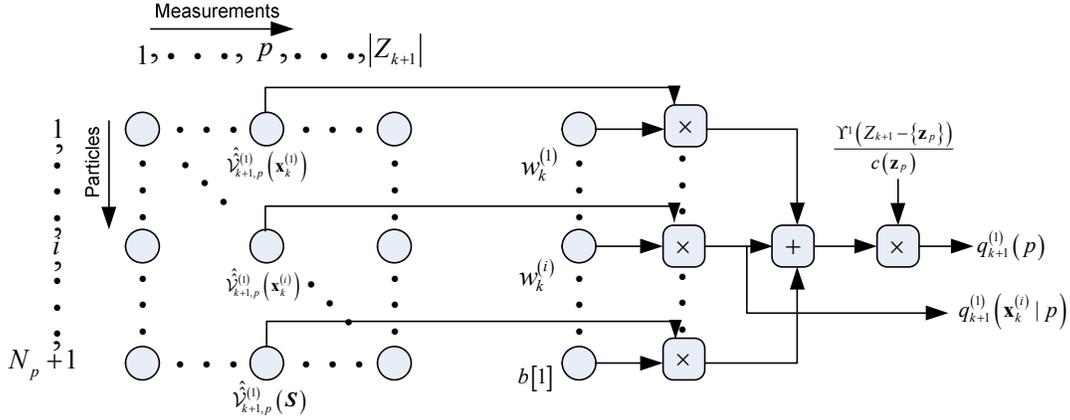

Figure 2. The schematic of how the proposal distributions $q_{k+1}^{(1)}(\mathbf{x}_k^{(i)} \mid p)$ and $q_{k+1}^{(1)}(p)$ are built according to (39) in order to pick up an appropriate existing particle $\mathbf{x}_k^{(i^n)}$ and a new measurement with index $p$ which is likely to be generated from a true target.

$$\bar{\mathbf{x}}_{k+1}^{(i)-} = \sum_{j=0}^{2L} W_j^{(m)} \cdot \boldsymbol{\chi}_{k+1|k}^{x(i)}(:,j), \quad i=1,\dots,N_p+1 \qquad (44)$$

$$\mathbf{P}_{k+1}^{(i)-} = \sum_{j=0}^{2L} W_j^{(c)} \cdot \left[ \boldsymbol{\chi}_{k+1|k}^{x(i)}(:,j) - \bar{\mathbf{x}}_{k+1}^{(i)-} \right] \cdot \left[ \boldsymbol{\chi}_{k+1|k}^{x(i)}(:,j) - \bar{\mathbf{x}}_{k+1}^{(i)-} \right]'$$

$$\boldsymbol{\mathcal{Y}}_{k+1|k}^{(i)}(:,j) = \begin{cases} h\left(\boldsymbol{\chi}_{k+1|k}^{x(i)}(:,j)\right) + \boldsymbol{\chi}_k^{a(i)}(9:10,j), \\ \qquad j=0,\dots,2L, \quad i=1,\dots,N_p \\ h\left(\boldsymbol{\chi}_{k+1|k}^{x(i)}(:,j)\right) + \boldsymbol{\chi}_k^{a(i)}(11:12,j), \\ \qquad j=0,\dots,2L, \quad i=N_p+1 \end{cases}$$

$$\bar{\mathbf{y}}_{k+1}^{(i)-} = \sum_{j=0}^{2L} W_j^{(m)} \cdot \boldsymbol{\mathcal{Y}}_{k+1|k}^{(i)}(:,j)$$

where, according to (8), the function $h(\cdot)$ is defined as

$$h\left(\boldsymbol{\chi}_{k+1|k}^{x(i)}(:,j)\right) = \left[ \sqrt{h\left(\boldsymbol{\chi}_{k+1|k}^{x(i)}(1,j)\right)^2 + h\left(\boldsymbol{\chi}_{k+1|k}^{x(i)}(3,j)\right)^2} \right.$$
$$\left. , \arctan\left(\frac{h\left(\boldsymbol{\chi}_{k+1|k}^{x(i)}(3,j)\right)}{h\left(\boldsymbol{\chi}_{k+1|k}^{x(i)}(1,j)\right)}\right) \right]' \qquad (45)$$

The equations of the measurement update are as follows

$$\mathbf{P}_{\mathbf{y}_{k+1}\mathbf{y}_{k+1}} = \sum_{j=0}^{2L} W_j^{(c)} \cdot \left[ \boldsymbol{\mathcal{Y}}_{k+1|k}^{(i)}(:,j) - \bar{\mathbf{y}}_{k+1}^- \right]$$
$$\times \left[ \boldsymbol{\mathcal{Y}}_{k+1|k}^{(i)}(:,j) - \bar{\mathbf{y}}_{k+1}^- \right]'$$
$$\mathbf{P}_{\mathbf{x}_{k+1}\mathbf{y}_{k+1}} = \sum_{j=0}^{2L} W_j^{(c)} \cdot \left[ \boldsymbol{\chi}_{k+1|k}^{x(i)}(:,j) - \bar{\mathbf{x}}_{k+1}^{(i)-} \right] \qquad (46)$$
$$\times \left[ \boldsymbol{\mathcal{Y}}_{k+1|k}^{(i)}(:,j) - \bar{\mathbf{y}}_{k+1}^- \right]'$$

$$\mathcal{K} = \mathbf{P}_{\mathbf{x}_{k+1}\mathbf{y}_{k+1}} \cdot \mathbf{P}_{\mathbf{y}_{k+1}\mathbf{y}_{k+1}}^{-1}$$
$$\bar{\mathbf{x}}_{k+1}^{(i)} = \bar{\mathbf{x}}_{k+1}^{(i)-} + \mathcal{K}\left(\mathbf{z}_{p_{k+1}} - \bar{\mathbf{y}}_{k+1}^-\right)$$
$$\mathbf{P}_{k+1}^{(i)} = \mathbf{P}_{k+1}^{(i)-} - \mathcal{K}\left(\mathbf{P}_{\mathbf{y}_{k+1}\mathbf{y}_{k+1}}\right)\mathcal{K}'$$

where $\mathbf{z}_{p_{k+1}}$ is the measurement of the RFS $Z_{k+1}$ whose index is the auxiliary variable $p_{k+1}$.

Now, thanks to the by-products of the UT measurement update, we have sufficient means at our disposal to design an efficient proposal distribution $q_{k+1}^{(1)}(\mathbf{x}_{k+1} \mid \mathbf{x}_k, p)$. We can generate the $n$th new sample, $\{\mathbf{x}_{k+1}^{d,(n)}\}_{n=1}^{N^{(1)}}$, according to the following Gaussian distribution:

$$q_{k+1}^{(1)}\left(\mathbf{x}_{k+1}^d \mid \mathbf{x}_k^{(n)}, p_{k+1}\right) = \mathcal{N}\left(\mathbf{x}_{k+1}^d; \bar{\mathbf{x}}_{k+1}^{(n)}, \mathbf{P}_{k+1}^{(n)}\right), \qquad (47)$$

where $\bar{\mathbf{x}}_{k+1}^{(i^n)}$ and $\mathbf{P}_{k+1}^{(i^n)}$ are given by (46).

In the case of undetected targets, the optimal choice for sampling $\{\mathbf{x}_{k+1}^{m,(n)}\}_{n=1}^{N^{(2)}}$ would be the extended single-state Markov transition density[2]:

$$q_{k+1}^{(2)}\left(\mathbf{x}_{k+1}^m \mid \mathbf{x}_k^{(n)}\right) = f^a\left(\mathbf{x}_{k+1}^m \mid \mathbf{x}_k^{(n)}\right), \quad n=1,\dots,N_p+1. \qquad (48)$$

The importance weights are computed to correct the discrepancies due to the usage of the first-stage weights, as it is necessary for the auxiliary particle filter [21]. Note that we should know the true values for functions $\Upsilon^0(Z_{k+1})$, $\Upsilon^1(Z_{k+1} - \{\mathbf{z}_p\})$ (for detected targets), and $\Upsilon^1(Z_{k+1})$ (for undetected targets) in order to compute the importance weights. In contrast to the predicted estimate, $\hat{D}_{k+1|k}[p_D L_{\mathbf{z}_p}]$, the update estimate of $D_{k+1|k}[p_D L_{\mathbf{z}_p}]$ which we will show with $\hat{D}_{k+1|k}[p_D L_{\mathbf{z}_p}]$, $p=1,\dots,|Z_{k+1}|$ can be obtained with new drawn samples $\{\mathbf{x}_{k+1}^{d,(n)}\}_{n=1}^{N^{(1)}}$ instead of the existing particles $\{\mathbf{x}_k^{(n)}\}_{n=1}^{N_p+1}$ as follows

---

[2] Indeed, discussed later on in this section, we will better understand the reason for its optimality after we show that the corresponding importance weights become uniform.

$$\hat{D}_{k+1|k}[p_D L_{z_p}] = \sum_{i=1}^{N_p+1} \int_E p_D(\mathbf{x}_{k+1}) \cdot L_{z_p}(\mathbf{x}_{k+1})$$

$$\times p_S^g(\mathbf{x}_k^{(i)}) \cdot \frac{f^a(\mathbf{x}_{k+1} \mid \mathbf{x}_k^{(i)}) \cdot D_{k|k}^g(\mathbf{x}_k^{(i)})}{q_{k+1}^{(1)}(\mathbf{x}_{k+1}, \mathbf{x}_k^{(i)} \mid p)} \cdot \delta_{\mathbf{x}_k^{(i)}}(d\mathbf{x}_k)$$

$$\times q_{k+1}^{(1)}(d\mathbf{x}_{k+1} \mid \mathbf{x}_k^{(i)}, p)$$

$$= \frac{\sum_{i=1}^{N_p} \mathcal{V}_{k+1,p}^{(1)}(\mathbf{x}_k^{(i)}) \cdot w_k^{(i)} + \mathcal{V}_{k+1,p}^{(1)}(\mathbf{S}) \cdot b[1]}{\left|\mathfrak{T}_{k+1,p}\right|} \tag{49}$$

$$\times \sum_{n \in \mathfrak{T}_{k+1,p}} \left[ \frac{p_D(\mathbf{x}_{k+1}^{d,(n)}) \cdot L_{z_p}(\mathbf{x}_{k+1}^{d,(n)}) \cdot p_S^g(\mathbf{x}_k^{(i^n)})}{q_{k+1}^{(1)}(\mathbf{x}_{k+1}^{d,(n)} \mid \mathbf{x}_k^{(i^n)}, p_{k+1}^{(n)})} \right.$$

$$\left. \times \frac{f^a(\mathbf{x}_{k+1}^{d,(n)} \mid \mathbf{x}_k^{(i^n)})}{\mathcal{V}_{k+1,p}^{(1)}(\mathbf{x}_k^{(i^n)})} \right],$$

where, if we consider new samples whose three-tuples have picked up the $p$th new measurement, then $\mathfrak{T}_{k+1,p}$ is the set of the indices of those new samples:

$$\mathfrak{T}_{k+1,p} = \left\{ n : p_{k+1}^{(n)} = p \right\}. \tag{50}$$

As a result, we can compute the update values (almost true values) of the functions $\Upsilon^0(Z_{k+1})$, $\Upsilon^1(Z_{k+1} - \{\mathbf{z}_p\})$, and $\Upsilon^1(Z_{k+1})$. We replace $D_{k+1|k}[p_D L_{z_p}]$ with $\hat{D}_{k+1|k}[p_D L_{z_p}]$ to obtain $\hat{\Upsilon}^0(Z_{k+1})$, $\hat{\Upsilon}^1(Z_{k+1} - \{\mathbf{z}_p\})$, and $\hat{\Upsilon}^1(Z_{k+1})$. The importance weights of the detected targets are computed as

$$w_{k+1}^{(1)}\left(\mathbf{x}_{k+1}^{d,(n)}, \mathbf{x}_k^{(i^n)}, p_{k+1}^{(n)}\right) = \frac{1}{N^{(1)}} \cdot p_D(\mathbf{x}_{k+1}^{d,(n)}) \cdot L_{z_{p_{k+1}^{(n)}}}(\mathbf{x}_{k+1}^{d,(n)})$$

$$\times \frac{1}{c\left(\mathbf{z}_{p_{k+1}^{(n)}}\right)} \cdot \frac{\hat{\Upsilon}^1\left(Z_{k+1} - \left\{\mathbf{z}_{p_{k+1}^{(n)}}\right\}\right)}{\hat{\Upsilon}^0(Z_{k+1})} \cdot f^a(\mathbf{x}_{k+1}^{d,(n)} \mid \mathbf{x}_k^{(i^n)}) \tag{51}$$

$$\times \frac{p_S^g(\mathbf{x}_k^{(i^n)}) \cdot D_{k|k}^g(\mathbf{x}_k^{(i^n)})}{q_{k+1}^{(1)}\left(\mathbf{x}_{k+1}^{d,(n)}, \mathbf{x}_k^{(i^n)}, p_{k+1}^{(n)}\right)}, \quad n = 1, \dots, N^{(1)}$$

$$, \quad i^n \in \{1, \dots, N_p\}.$$

*Remark 1:* Consider $\mathbf{x}_k^{(i^n)} \neq \mathbf{S}$ in (51) so that $f^a(\mathbf{x}_{k+1}^{d,(n)} \mid \mathbf{x}_k^{(i^n)}) = f(\mathbf{x}_{k+1}^{d,(n)} \mid \mathbf{x}_k^{(i^n)})$ equals the Gaussian density $\mathcal{N}(\mathbf{x}_{k+1}^{d,(n)}; F(\mathbf{x}_k^{(i^n)}(5)) \cdot \mathbf{x}_k^{(i^n)}, \mathbf{G} \cdot \mathbf{Q} \cdot \mathbf{G}')$. As it is evident, $(\mathbf{G}_{5\times3} \cdot \mathbf{Q}_{3\times3} \cdot \mathbf{G}')_{5\times5}$ is not a full rank covariance matrix (two out of five eigenvalues are zero) and therefore the Gaussian $f(\mathbf{x}_{k+1}^{d,(n)} \mid \mathbf{x}_k^{(i^n)})$ cannot be evaluated at a given $\mathbf{x}_{k+1}^{d,(n)}$. Instead of $f(\mathbf{x}_{k+1}^{d,(n)} \mid \mathbf{x}_k^{(i^n)})$, we may evaluate the below full rank Gaussian density

$$\mathcal{N}\left((\mathbf{G}' \cdot \mathbf{G})^{-1} \cdot \mathbf{G}' \cdot \left(\mathbf{x}_{k+1}^d - F(\mathbf{x}_k^{(n)}(5)) \cdot \mathbf{x}_k^{(n)}\right); \mathbf{0}_{3\times1}, \mathbf{Q}\right). \tag{52}$$

The importance weights of undetected targets are computed as

$$w_{k+1}^{(2)}\left(\mathbf{x}_{k+1}^{m,(n)}, \mathbf{x}_k^{(i^n)}\right) = \frac{(1 - p_D)}{N^{(2)}} \cdot \frac{\hat{\Upsilon}^1(Z_{k+1})}{\hat{\Upsilon}^0(Z_{k+1})} \tag{53}$$

$$\times \frac{f^a\left(\mathbf{x}_{k+1}^{m,(n)} \mid \mathbf{x}_k^{(i^n)}\right) \cdot p_S^g(\mathbf{x}_k^{(i^n)}) \cdot D_{k|k}^g(\mathbf{x}_k^{(i^n)})}{q_{k+1}^{(2)}\left(\mathbf{x}_{k+1}^{m,(n)}, \mathbf{x}_k^{(i^n)}\right)}.$$

$$n = 1, \dots, N^{(2)}, \quad i^n \in \{1, \dots, N_p\}$$

If we consider (43), importance weights of undetected targets is transformed into the following form:

$$w_{k+1}^{(2)}\left(\mathbf{x}_{k+1}^{m,(n)}, \mathbf{x}_k^{(i^n)}\right) = \frac{\sum_{i=1}^{N_p} p_S \cdot w_k^{(i)} + b[1]}{N^{(2)}} \cdot \frac{\hat{\Upsilon}^0(Z_{k+1})}{\hat{\Upsilon}^1(Z_{k+1})} \tag{54}$$

$$n = 1, \dots, N^{(2)}, \quad i^n \in \{1, \dots, N_p\}.$$

We can see from (54) that the importance weights of undetected targets are uniform. This results in the minimum importance weight variance (equal to zero).

The expected number of targets at the time-step $k+1$ in the observation area is computed by taking summation of $N_p$ elements of the concatenated set $\{w_{k+1}^{(1)}(.,.,.)\}_{n=1}^{N^{(1)}} \bigcup \{w_{k+1}^{(2)}(.,.)\}_{m=1}^{N^{(2)}}$.

*Remark 2:* It is not required to perform resampling at the end of each cycle of Auxiliary particle filter family. In fact, resampling step is done before computing importance weights. While we are picking up auxiliary variables, we are doing resampling.

The process of applying UT to implement ACPHD is illustrated in the schematic in Fig. 3.

## V. SIMULATIONS

We demonstrate the superiority of the proposed Unscented-Auxiliary CPHD (U-ACPHD) filter over the SMC-PHD, SMC-CPHD and Unscented-Auxiliary PHD (U-APHD) filter by simulation results. The implementation method for the SMC-PHD and SMC-CPHD filter are described in [12,13]. We follow the same procedure described for the U-ACPHD filter in order to implement the U-APHD filter expect that the update and correction steps of the U-APHD filter is modified according to the PHD recursion [15].

In our simulations, we confine observations to a square region with sides equal to 1km while the sensor is located in $(0,0)$. The simulation scenario parameters are denoted in Table 1.

For the SMC-PHD and SMC-CPHD filter, we set the numbers of particles assigned to sample from the intensity function of newly born and persistent targets to 500 and 2500 respectively, which are fixed regardless of the expected number of targets. For the birth intensity function, we set

$$\mathbf{m}_b = [500, 0, 500, 0, 0]' \text{ and } \mathbf{Q}_b = diag\left(15^2, 5^2, 15^2, 5^2, 0.1^2\right).$$

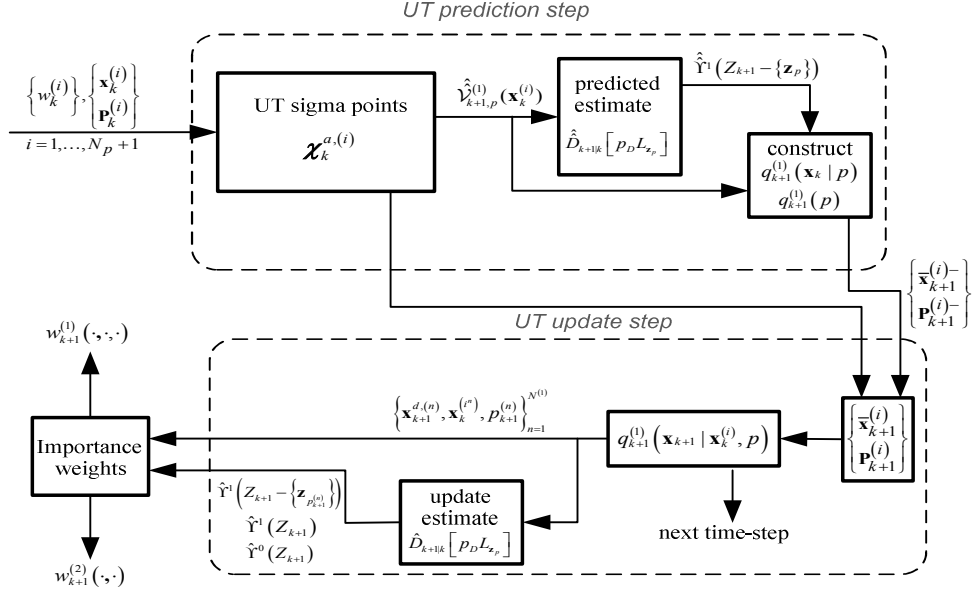

Figure 3. The schematic of how UT can be used to implement ACPHD filter and construct proposal distributions as well as computing importance weights.

TABLE 1. SIMULATION SCENARIO PARAMETERS

| Parameter | Value |
|---|---|
| $T$ | 1 s |
| $\sigma_\varepsilon$ | 0.1 |
| $\sigma_\omega$ | $\pi/180$ rad/s |
| $\sigma_\theta$ | $0.5 \times \pi/180$ rad |
| $\sigma_r$ | 1 m |
| $p_D$ | 0.95 |
| $p_S$ | 0.99 |
| $N^{(1)}$ | 2500 |
| $N^{(2)}$ | 500 |
| Simulation length | 90 time-steps |

There are 5 targets appearing and disappearing during the simulation time of 90 time-steps (nearly 100 time-steps), so the predicted number of newly born targets $b[1] = \gamma_b$ is 5/100. The spatial distribution $c(\mathbf{z})$ is uniform over range and bearing domain $[0,1000] \times [0, \pi/2]$ and average number of clutter points $\lambda$ equals 10. We set $\gamma$ and $\beta - \alpha^2$ to 2 and 0 respectively in (34).

Our scenario of interest, consisting of five targets whose trajectories are merged into clutter, is shown in Fig. 4. In addition, the related initial states and appearance and disappearance time-steps of targets used in the simulation scenario are all illustrated in Table 2. The true number of targets at each time-step can be obtained from the target appearance and disappearance times in Table 2.

TABLE 2. TARGETS' INITIAL STATES AND THEIR APPEARANCE AND DISAPPEARANCE TIME-STEPS

| TARGET NUMBER | INITIAL STATE (M,M/S,RAD/S) | APPEARANCE TIME (TIME-STEP) | DISAPPEARANCE TIME (TIME-STEP) |
|---|---|---|---|
| 1 | $[505,-5,490,-5,0]'$ | 1 | 70 |
| 2 | $[485,5,525,-5,0]'$ | 5 | 74 |
| 3 | $[505,5,505,-5,0]'$ | 11 | 80 |
| 4 | $[495,5,490,5,0]'$ | 15 | 84 |
| 5 | $[500,-5,510,5,0]'$ | 21 | 90(END) |

We computed the OSPA, cardinality, and localization distance between the set of target state position estimates and true target positions at each time-step in order to compare performance of the four aforementioned tracking algorithms (see [23] for more details about these distances). Each distance is obtained by averaging over 100 Monte Carlo runs. Two parameters of the OSPA distance, the OSPA order $(\rho)$ and the OSPA cut-off $(c)$ are set to 2 and 150 m respectively. The SMC-PHD, SMC-CPHD, U-APHD, and U-ACPHD filter approximates intensity function with particle samples and do not explicitly provide any state estimate. We apply the natural clustering methods for extracting target states from sample representation of the U-APHD and U-ACPHD filter. The method is valid thanks to the principle of auxiliary variables and described in [14,15]. However, it would not work for the case of the SMC-PHD and SMC-CPHD filters and we have to resort to the Kmeans algorithm implemented in MATLAB, where we set its parameters 'distance' and 'replicates' to 'city' and '5' respectively.

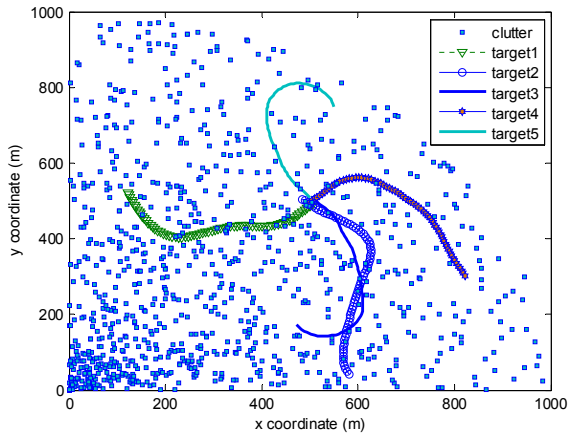

Figure 4. Trajectories of true targets emerged in clutters (average number of Clutters is 10). The simulation totally includes 90 time-steps.

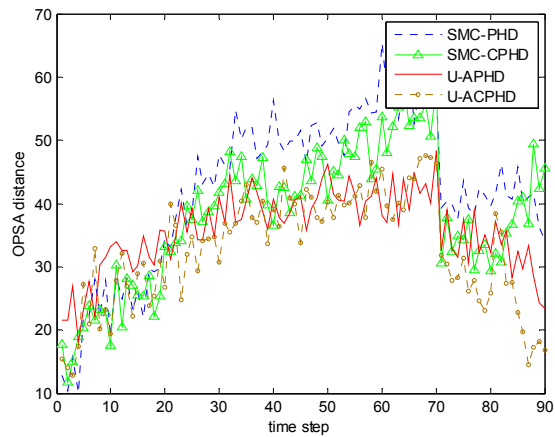

Figure 5. The OSPA distance for different filters versus simulation time-step, averaged over 100 Monte Carlo runs. An average number of Clutter points is 10.

As shown in Fig. 5, the OSPA distance is getting larger when we see a growth of the number of targets much like what happens for the period started from the time-step 21 up to the time-step 70, common for all filters. This is due to using fixed number of particles regardless of the expected number of targets. As a consequence of this strategy, the number of particles assigned to each tracked target decreases and this leads to poor estimation performance.

Apart from this fact, as shown in Figs. 5-7, the U-ACPHD outperforms in terms of the localization accuracy and the cardinality estimation because the U-ACPHD filter uses two auxiliary variables to utilize the most current information to improve both the cardinality density and intensity function. Of the other methods tested, the U-APHD filter offers the next best performance, and that is because it applies auxiliary variables to only improve the estimation of intensity function. We can see that the SMC-PHD filter has the worst results in cardinality estimation and localization performance compared to other filters.

Indeed, as pointed out in [5], the value of the expected number of target for the SMC-PHD filter is very unstable in the presence of misdetection and there is a high probability to lose a confirmed track. This justifies its higher OSPA, cardinality and localization distances for the last time-steps.

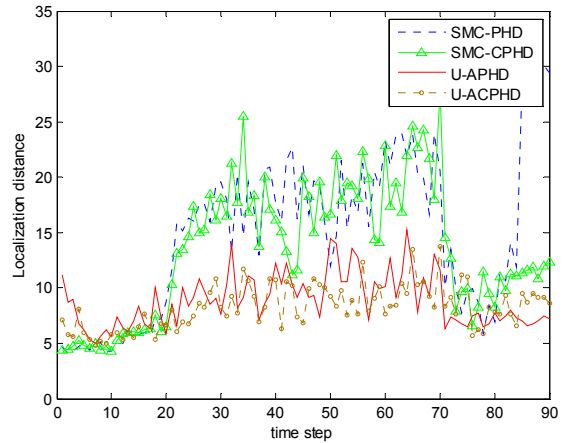

Figure 6. The localization distance averaged over 100 Monte Carlo runs. An average number of Clutter points is 10. Compared to the auxiliary particle based filters, both the SMC filters have failed in performance of localization estimation fo time-steps in which all the targets are present.

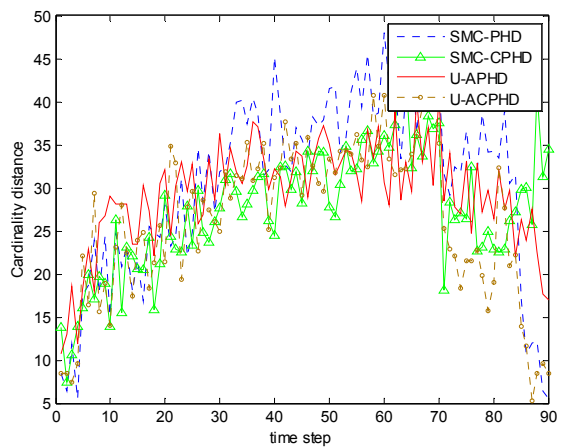

Figure 7. The cardinality distance averaged over 100 Monte Carlo runs. An average number of Clutter points is 10. Apart from the SMC-PHD filter, other filters have nearly similar performance.

In order to assess the impact of average number of clutter points on estimation of intensity function, we compute the OSPA, cardinality, and localization distance over the 100 Monte Carlo runs, for various values of $\lambda$. These distances are then averaged over the entire period of simulation time (90 time-steps). The results are represented in Table 3.

According to the Table 3, the U-ACPHD filter again outperforms the rest of filters in terms of all average distances for all tested values of $\lambda$. Its robustness and accuracy in different environments (especially in high-clutter environments) justify the complexity involved in the implementation of the U-ACPHD filter. To our surprise, the SMC-PHD filter has the second best performance. One part of the reason for this behavior is traced to the fact that the CPHD filter is more robust than the PHD filter from increase in the average clutter rate especially near the origin.

TABLE 3
THE AVERAGE OSPA PERFORMANCE OF DIFFERENT FILTERS FOR VARIOUS
VALUES OF THE AVERAGE NUMBER OF CLUTTER POINTS $\lambda$.

| $\lambda = 10$ | SMC-PHD FILTER | SMC-CPHD FILTER | U-APHD FILTER | U-ACPHD FILTER |
|---|---|---|---|---|
| OSPA DIS. | 42.4468 | 38.3107 | 35.8004 | 32.7478 |
| LOC. DIS. | 14.9869 | 13.5236 | 8.8654 | 8.2779 |
| CAR. DIS. | 30.5740 | 27.5587 | 29.2819 | 26.3616 |
| $\lambda = 30$ | SMC-PHD FILTER | SMC-CPHD FILTER | U-APHD FILTER | U-ACPHD FILTER |
| OSPA DIS. | 51.2200 | 54.7985 | 64.5026 | 46.6291 |
| LOC. DIS. | 17.3767 | 19.0280 | 11.5215 | 13.2210 |
| CAR. DIS. | 38.7589 | 41.1676 | 58.1376 | 36.6109 |
| $\lambda = 50$ | SMC-PHD FILTER | SMC-CPHD FILTER | U-APHD FILTER | U-ACPHD FILTER |
| OSPA DIS. | 63.7661 | 65.9517 | 87.8809 | 61.3166 |
| LOC. DIS. | 21.9895 | 24.3297 | 10.5030 | 22.3662 |
| CAR. DIS. | 49.3045 | 49.6968 | 84.0311 | 44.4617 |

TABLE 4.
THE AVERAGE MEAN AND AVERAGE STANDARD DEVIATION OF RUN TIMES PER
TIME-STEP FOR $\lambda = 10, 30$, AND 50. THOSE EXPERIMENT-FILTER PAIRS WHOSE
METHODS ARE NOT REAL-TIME ARE SHOWN IN BOLDFACE.

| EXPERIMENT | SMC-PHD | | SMC-CPHD | | U-APHD | | U-ACPHD | |
|---|---|---|---|---|---|---|---|---|
| | MEAN | SD | MEAN | SD | MEAN | SD | MEAN | SD |
| $\lambda = 10$ | 0.1065 (s) | 0.0639 (s) | 0.6912 (s) | 0.2724 (s) | **2.2044 (s)** | 0.3184 (s) | **3.0445 (s)** | 0.9568 (s) |
| $\lambda = 30$ | 0.1310 (s) | 0.0565 (s) | 0.6338 (s) | 0.0631 (s) | **2.8268 (s)** | 0.0569 (s) | **3.6474 (s)** | 0.0503 (s) |
| $\lambda = 50$ | 0.1451 (s) | 0.0544 (s) | 0.9490 (s) | 0.0667 (s) | **3.7033 (s)** | 0.0734 (s) | **6.6186 (s)** | 2.0743 (s) |

Near the origin, because of the higher incidence of false alarms, false tracks can be considered as candidates of targets with very low probability of detection. However, the SMC-PHD is not good to keep track of targets with low probability of detection and seems to be less vulnerable to higher clutter ratings.

The reason for the superiority of the SMC-PHD filter over the U-APHD filter in scenarios with high clutter lies with the usage of auxiliary variables which improves the localization estimation accuracy for true tracks and leads to low localization distances. As a result, the APHD filter accounts for false tracks and tries to prolong them. The consequences of these tendencies are little localization distance versus high cardinality distance. The solution to these vulnerabilities is to enjoy the benefits of synergy obtained by combination of updating cardinality distribution as well as using auxiliary variables to estimate the intensity function. This is what exactly the U-ACPHD filter is doing, which yields a real compromise between cardinality and localization distance in harsh environments , as it is evident in Table 3.

Another comparison metric we are using is the computational efficiency of these algorithms.

If we denote the number of current measurements by $|Z_{k+1}|$ and the number of particles by $N_p$, the SMC-PHD and SMC-CPHD filter have computational complexity $\mathcal{O}(|Z_{k+1}| \cdot N_p)$ and $\mathcal{O}(|Z_{k+1}|^2 \cdot \log^2 |Z_{k+1}| \cdot N_p)$ respectively. The two other filters with auxiliary particle implementation clearly demand more computation resources since they need to obtain $2L+1$ UT sigma points for every particle.

To compare the four aforementioned filters in terms of run times, we compute mean and standard deviation of CPU run times per time-step on a Core i5-3570K Ivy Bridge 3.4GHz, where each result is averaged over 100 Monte Carlo runs. The results are summarized in Table 4.

We also study the average (out of 100 Monte Carlo runs) number of picking the newly born sample **S** by both the U-APHD and U-ACPHD filter. The results are shown in Fig. 8.

For example, at the time of first target appearance, time-step 5, the $N^{(1)} + N^{(2)} + 1$ th particle, **S**, is selected nearly 200 times by the U-ACPHD filter out of $N^{(1)} + N^{(2)} + 1$ existing particles to form $N^{(1)}$ three-tuples $\{\mathbf{x}_{k+1}^{d,(n)}, \mathbf{x}_k^{(i*)}, p_{k+1}^{(n)}\}_{n=1}^{N^{(1)}}$. This shows that the U-ACPHD filter is sensitive to birth events. For other time-steps at which there is no birth event, the repetition of picking the existing particle **S** decreases. According to Fig. 8, the U-ACPHD filter has better sensitivity to birth event than the U-APHD filter thanks to more information available by updating the cardinality statistics. Furthermore, the U-ACPHD filter usually has less repetition of the newly born sample when there is no birth. This fact could be verified by Fig. 8 when it shows that the U-ACPHD filter contains more discontinuous points, which are indicative that the particle **S** is not selected at those points.

## VI. CONCLUSION

In this paper, we have derived an auxiliary particle implementation of the CPHD filter based on the UT algorithm. Our proposed algorithm exploits the UT algorithm to construct the proposal distributions required to draw a set of auxiliary variables containing measurement indices and existing particles. This works to boost the performance in approximating intensity function just like how auxiliary particle filter samples on a higher dimensional space and achieves more accurate estimation than those achieved with Sampling-importance resampling (SIR) scheme.

We have overcome the difficulty of using auxiliary particle for implementation of the highly complex CPHD filter recursion with defining potential functions which help with approximation of the elementary symmetric functions. To that end, we have discussed in detail how to apply UT to compute these potential functions before and after generating current samples.

We have compared our proposed algorithms against the SMC-PHD and the SMC-CPHD and the U-APHD filter based on different metrics such as cardinality and localization distance, algorithm run times, and sensitivity to new target appearance. As it is evident in Table 3, numerical results have shown our proposed algorithm superiority over its counterparts especially in scenarios with high clutter, although the U-ACPHD filter takes up more computation time because of the UT steps and more complex recursion.

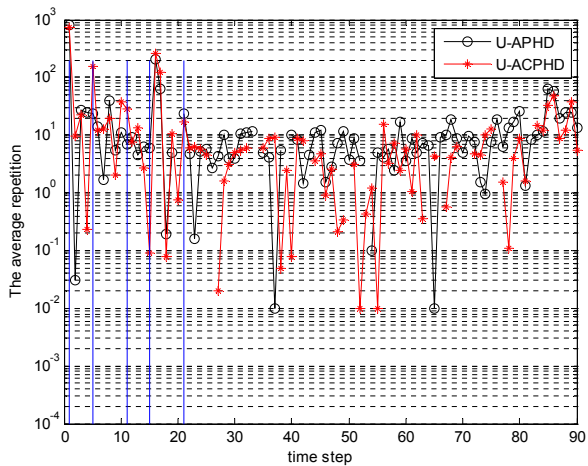

Figure 8. Repetition of the $N^{(1)} + N^{(2)} + 1$ th particle, $\mathbf{S}$, as the selected existing particle, averaged over 100 Monte Carlo runs. Blue lines indicate time-steps at which birth occurs. Any discontinuity demonstrates that there is no selection of the particle, $\mathbf{S}$, out of $N^{(1)}$ selected existing particles.